\pdfoutput=1
\documentclass[conference]{IEEEtran}
\IEEEoverridecommandlockouts
\usepackage{mathtools,amssymb,amsthm,amsfonts,nicefrac,xfrac}
\usepackage{graphicx}
\usepackage{textcomp}
\makeatletter
\let\MYcaption\@makecaption
\makeatother
\usepackage{subcaption}
\makeatletter
\let\@makecaption\MYcaption
\makeatother
\usepackage{csquotes}
\usepackage{xcolor}
\usepackage{tikz}
\usepackage[backend=biber,style=ieee,sortlocale=en-US,url=true,doi=false,date=year,minnames=2,maxnames=4,isbn=false, sortcites, dashed=false]{biblatex}
\usepackage{microtype}
\usepackage{hyperref}

\usetikzlibrary{positioning,calc,fit,backgrounds,matrix,shapes,arrows.meta,graphs,quotes,zx-calculus}

\newcommand*{\ket}[1]{\ensuremath{|#1\rangle}}

\DeclarePairedDelimiter\abs{\lvert}{\rvert}

\newtheorem{example}{Example}

\hypersetup{
	pdftitle={The Basis of Design Tools for Quantum Computing: Arrays, Decision Diagrams, Tensor Networks, and ZX-Calculus},
	pdfsubject={DAC 2022},
	pdfauthor={Robert Wille, Lukas Burgholzer, Stefan Hillmich, Thomas Grurl, Alexander Ploier, Tom Peham}
}

\AtEveryBibitem{% Clean up the bibtex rather than editing it
  \clearname{editor}%
  \clearfield{isbn}%
  \clearfield{issn}%
  \clearfield{pages}%
}

\tikzset{%
	font={\footnotesize},
	vertex/.style={draw,circle,inner sep=0pt,minimum width=0.5cm,minimum height=0.5cm,font=\small, scale=0.9},
	terminal/.style={draw,regular polygon,regular polygon sides=4,inner sep=0pt,minimum width=0.5cm,minimum height=0.5cm,font=\small, scale=1.0},
	zeroterm/.style={below,inner sep=0pt,font=\scriptsize, scale=0.9}
}

\begin{document}

\title{\Huge The Basis of Design Tools for Quantum Computing: \\ {\LARGE Arrays, Decision Diagrams, Tensor Networks, and ZX-Calculus}}
\IEEEspecialpapernotice{(Invited Paper)}

\author{\IEEEauthorblockN{Robert Wille\textsuperscript{1,2*}, Lukas Burgholzer\textsuperscript{3*}, Stefan Hillmich\textsuperscript{3*}, Thomas Grurl\textsuperscript{3}, Alexander Ploier\textsuperscript{3}, Tom Peham\textsuperscript{3}}
	\IEEEauthorblockA{\textsuperscript{1} Chair for Design Automation, Technical University of Munich, Germany}
	\IEEEauthorblockA{\textsuperscript{2} Software Competence Center Hagenberg (SCCH) GmbH, Austria}
	\IEEEauthorblockA{\textsuperscript{3} Institute for Integrated Circuits, Johannes Kepler University Linz, Austria}
	\IEEEauthorblockA{\textsuperscript{*}Corresponding Authors: \href{mailto:robert.wille@tum.de}{robert.wille@tum.de}, \href{mailto:lukas.burgholzer@jku.at}{lukas.burgholzer@jku.at}, \href{mailto:stefan.hillmich@jku.at}{stefan.hillmich@jku.at}\\
	\url{https://www.cda.cit.tum.de/research/quantum/}\vspace*{-3mm}}
}

\maketitle

\begin{abstract}
Quantum computers promise to efficiently solve important problems classical computers never will. However, in order to capitalize on these prospects, a fully automated quantum software stack needs to be developed. This involves a multitude of complex tasks from the classical simulation of quantum circuits, over their compilation to specific devices, to the verification of the circuits to be executed as well as the obtained results. All of these tasks are highly non-trivial and necessitate efficient \mbox{data structures} to tackle the inherent complexity. Starting from rather straight-forward arrays over decision diagrams (inspired by the design automation community) to tensor networks and the \mbox{ZX-calculus}, various complementary approaches have been proposed. This work provides a look \enquote{under the hood} of today’s tools and showcases how these means are utilized in them, e.g., for simulation, compilation, and verification of quantum circuits.
\end{abstract}

\section{Introduction}

We are at the dawn of a new computing age in which quantum computers will find their way into practical applications such as cryptography~\cite{shorPolynomialtimeAlgorithmsPrime1997}, chemistry~\cite{kandalaHardwareefficientVariationalQuantum2017}, medicine~\cite{cordierBiologyMedicineLandscape2021}, physics~\cite{doi:10.1126/science.abe8770}, finance~\cite{hermanSurveyQuantumComputing2022}, and machine learning~\cite{huangProvablyEfficientMachine2022}.
In many instances, quantum computing is believed to provide efficient solutions for problems which are out of reach for classical computers.
Besides the ongoing discovery of new potential applications, the capabilities of currently available quantum computers are rapidly improving as, e.g., witnessed by IBM's ambitious road map for scaling quantum technology to more than 1000 qubits by 2023~\cite{gambettaIBMRoadmapScaling2020}.

Due to an increased number of qubits with increased coherence time as well as faster operations with higher fidelity, increasingly large quantum circuits can reliably be executed on actual devices.
With this increase in computational power comes the need for corresponding software solutions and tools that aid users and developers in making best use of the available hardware.
Similar to the design of classical circuits and systems, realizing conceptual quantum algorithms on actual devices requires a multitude of complex design tasks.
Some of the most important tasks are:
\begin{itemize}
    \item \emph{Classical simulation}: Simulating the execution of a quantum circuit on classical computers is an extremely important task in the development and testing of new applications and use cases. In addition to lower costs, it offers detailed insights on the quantum state during the execution of a quantum circuit that is physically unavailable when running the circuit on an actual quantum computer~\cite{VMP:2009,zulehnerAdvancedSimulationQuantum2019,jonesQuESTHighPerformance2018,PhysRevLett.116.250501,hillmichAccurateNeededEfficient2020,grurlnoiseaware2022}.
    \item \emph{Compilation}: Similar to classical circuits and systems, quantum circuits are initially described at a rather high abstraction level and need to be compiled to a representation that adheres to all the constraints imposed by the target device (e.g., limited gate-set and/or limited connectivity)~\cite{boteaComplexityQuantumCircuit2018,zulehnerEfficientMethodologyMapping2019,hanerSoftwareMethodologyCompiling2018,smithQuantumComputationalCompiler2019,liTacklingQubitMapping2019}.
    \item \emph{Verification}: Since compilation significantly changes the structure of quantum circuits, it is crucial to ensure that the resulting circuits still realize the originally intended functionality. To this end, verification (or, more precisely, equivalence checking) methods are employed to guarantee equivalence~\cite{yamashitaFastEquivalencecheckingQuantum2010, burgholzerAdvancedEquivalenceChecking2021,viamontesCheckingEquivalenceQuantum2007,niemannEquivalenceCheckingMultilevel2014,wangXQDDbasedVerificationMethod2008,smithQuantumComputationalCompiler2019, amyLargescaleFunctionalVerification2019, hongTensorNetworkBased2020}.
\end{itemize}

Either due to the inherent exponential size of the underlying representations of quantum states and operations or the huge amount of degrees of freedom, each of these design tasks represents a computationally hard challenge.
Consequently, efficient data structures and methods are needed to tackle these challenges.
In this work, we provide a brief overview of various complementary data structures that have been proposed in the past and briefly discuss how each of them has been used to efficiently solve the above mentioned design tasks.
With this, we hope to provide the interested reader with an intuition on the different kinds of approaches available and the necessary pointers to dive deeper into the wide range of possible methods and solutions.

The rest of this work is structured as follows:
\autoref{sec:arrays} reviews the basics of quantum computing and shows how quantum states and operations are represented as one- and \mbox{two-dimensional} arrays in a straight-forward, yet hardly efficient fashion.
\autoref{sec:dds} introduces decision diagrams which enable representing quantum states and functionality in a more compact fashion in many cases by exploiting redundancies in the underlying representations.
\autoref{sec:tns} covers the basics of tensor networks which, instead of capitalizing on redundancies in the underlying representations, take advantage of the topological structure of certain quantum states and algorithms.
\autoref{sec:zx} demonstrates how the ZX-calculus---a graphical notation for quantum circuits equipped with a powerful set of rewrite rules---enables diagrammatic reasoning about quantum computing.
Finally, \autoref{sec:conclusion} concludes the paper.

\section{Arrays}\label{sec:arrays}

In quantum computing, vectors and matrices are often considered to be the most intuitive data structure for representing quantum objects. These structures can be directly realized using arrays and can be used for design automation tasks.
Here, we introduce this data structure along with a brief introduction to quantum computing.
The interested reader can find an in-depth introduction in~\cite{nielsenQuantumComputationQuantum2010}.

Similar to classical bits, \emph{quantum bits} (\emph{qubits}) can assume the states $0$ or $1$. These are are called computational basis states and---using Dirac notation---written as $\ket{0}$ and $\ket{1}$. Additionally, they can also assume an (almost) arbitrary linear combination (i.e.,~a~\emph{superposition}) of these two basis states. 
More precisely, the state of a qubit~$\ket{\psi}$ is given by \mbox{$\ket{\psi} = \alpha_0 \cdot \ket{0} + \alpha_1 \cdot \ket{1}$}, with $\alpha_0, \alpha_1 \in \mathbb{C}$ such that \mbox{$\abs{\alpha_0}^2 + \abs{\alpha_1}^2 = 1$}. 
The two factors $\alpha_0$ and $\alpha_1$ are the \emph{amplitudes} and denote how much the qubit is related to each of the two basis states. 
Measuring a qubit returns 0 with probability $\abs{\alpha_0}^2$ and 1 with probability $\abs{\alpha_1}^2$, respectively.
The individual amplitudes in a qubit are fundamentally not observable and measurements are the only way to extract information out of a qubit.

The concepts of a single qubit can be generalized to describe states composed of multiple qubits---commonly referred to as \emph{quantum registers}. 
An \mbox{$n$~qubit} register can assume $2^{n}$ basis states and is described by amplitudes $\alpha_0, \alpha_1, \dots \alpha_{2^n-1}$, which must satisfy the normalization constraint $\sum_{i\in \{0,1\}^n} |\alpha_i|^2 = 1$. 
Quantum states are often shortened to state vectors containing only the amplitudes, e.g., 
$\begin{bsmallmatrix} \alpha_{00}&\alpha_{01}&\alpha_{10}&\alpha_{11}\end{bsmallmatrix}^\mathrm{T}$ for two qubits. 

Quantum states can be manipulated using quantum operations. 
Quantum operations are inherently reversible and are described by unitary matrices. 
They are applied to quantum states by matrix-vector multiplication. 
Important \mbox{single-qubit} operations include the $\textup{NOT}=\begin{bsmallmatrix}0&1\\1&0\end{bsmallmatrix}$ operation, which negates the state of a qubit, and the Hadamard operation $\textup{H}=\sfrac{1}{\sqrt{2}}\begin{bsmallmatrix*}[r]1&1\\1&-1\end{bsmallmatrix*}$, which transforms a qubit from a basis state into a superposition.
There are also multi-qubit operations. 
The most prominent two-qubit operation is the controlled-NOT operation (CNOT), which negates the state of its target qubit iff the control qubit is in state $\ket{1}$. 

\begin{example}
	\label{exa:strong-matrix_vector_simulation}
	Consider the quantum register $ \ket{\psi} $ composed of two qubits, which is in the state
	%\footnote{This state can be constructed by applying a Hadamard operation to $ q_0 $ when the state is initialized to $ \ket{00} $.} 
	$\sfrac{1}{\sqrt{2}}\cdot\begin{bsmallmatrix}1&0&1&0\end{bsmallmatrix}^\mathrm{T}$.
	Applying a \textup{CNOT} operation with control on the first and target on the second qubit yields the output state~$\ket{\psi^\prime}$ determined by
	\begin{align*}
		\underbrace{\begin{bsmallmatrix}1&0&0&0\\0&1&0&0\\0&0&0&1\\0&0&1&0\end{bsmallmatrix}}_{\textup{CNOT}} 
		\: \cdot \: 
		\underbrace{\frac{1}{\sqrt{2}}\begin{bsmallmatrix}1\\0\\1\\0\end{bsmallmatrix}}_{\mathit{\ket{\psi}}} 
		\: = \: \underbrace{\frac{1}{\sqrt{2}}\begin{bsmallmatrix}1\\0\\0\\1\end{bsmallmatrix}}_{\mathit{\ket{\psi^\prime}}}.
	\end{align*}
	Measuring $\ket{\psi^\prime}$ (also known as \emph{Bell state}) collapses the state and returns $\ket{00}$ or $\ket{11}$, each with probability $\abs{\sfrac{1}{\sqrt{2}}}^2=\sfrac{1}{2}$.
\end{example}

The concepts reviewed above can be realized in a straightforward fashion: Vectors and matrices are described in terms of 1-dimensional and 2-dimensional arrays, respectively. While such a representation has huge potential for concurrent execution, it incurs a huge memory footprint, since the involved arrays growth exponentially with each considered qubit.
As a consequence, these memory requirements limit \mbox{array-based} simulation methods to rather small/moderate quantum computations (today's practical limit is less than 50 qubits~\cite{DeRaedt2019}).

\section{Decision Diagrams}\label{sec:dds}

\begin{figure}[t]
	\centering
	\begin{subfigure}[t]{0.4\linewidth}
		\centering
		\begin{tikzpicture}
		\matrix[matrix of math nodes, left delimiter={[},right delimiter={]}, inner xsep=0] (vector) {
			\frac{1}{\sqrt{2}}\\				
			0\\
			0\\
			\mathbf{\frac{1}{\sqrt{2}}}\\				
		};			
		\begin{scope}[on background layer, gray]	
		\node[right=0.6cm of vector-1-1.center] {$\ket{00}$};
		\node[right=0.6cm of vector-2-1.center] {$\ket{01}$};
		\node[right=0.6cm of vector-3-1.center] {$\ket{10}$};
		\node[right=0.6cm of vector-4-1.center] {$\ket{11}$};
		
		\draw[gray,-,dashed, very thick,shorten <= -0.6cm] ($(vector-2-1)!0.5!(vector-3-1)$) -- ++(-1.25,0) node[anchor=east] {\(q_1\)};
		
		\draw[gray,-,dashed, thick,shorten <= -0.6cm] ($(vector-1-1)!0.5!(vector-2-1)$) -- ++(-1,0) node[anchor=east] {\(q_0\)};
		\draw[gray,-,dashed, thick,shorten <= -0.6cm] ($(vector-3-1)!0.5!(vector-4-1)$) -- ++(-1,0) node[anchor=east] {\(q_0\)};
		
		\end{scope}
		\end{tikzpicture}
		\caption{Vector}
		\label{fig:strong-statevectorvector}
	\end{subfigure}\qquad
	\begin{subfigure}[t]{0.4\linewidth}
		\centering
		\begin{tikzpicture}	
		\matrix[matrix of nodes,ampersand replacement=\&,every node/.style=vertex,column sep={0.5cm,between origins},row sep={0.9cm,between origins}] (qmdd2) {
			\& \node (m1) {$q_1$}; \& \\
			\node (m2a) {$q_0$}; \& \& \node (m2b) {$q_0$}; \\
			\& \node[terminal] (t3) {1}; \& \\
		};
		
		\draw[thick] ($(m1)+(0,0.5cm)$) -- (m1) node[right, midway]{$\sfrac{1}{\sqrt{2}}$};
		
		\draw[thick] (m1.-135) -- ($(m1)!0.5!($(m2a)!0.5!(m2b)$) + (-5mm,0)$) -- (m2a.90);
		\draw (m1.-45) -- ($(m1)!0.5!($(m2a)!0.5!(m2b)$) + (5mm,0)$) -- (m2b.90);
		
		\draw[thick] (m2a.-135) -- ($(m2a.-135) - (1.2mm,2.0mm)$) -- (t3.135);
		\draw (m2a.-45) -- ($(m2a.-45)!0.2!(t3) + (0mm,0)$)  node[zeroterm] {$0$};;
		
		\draw (m2b.-135) -- ($(m2b.-135)!0.2!(t3) + (0mm,0)$)  node[zeroterm] {$0$};;
		\draw (m2b.-45) -- ($(m2b.-45) - (-1.2mm,2.0mm)$) -- (t3.45);		
		\end{tikzpicture}
		\caption{Decision diagram}
		\label{fig:strong-statevectordd}
	\end{subfigure}%
	\caption{Different representations of the Bell state}
	\label{fig:strong-statevector}
	\vspace{-1em}
\end{figure}
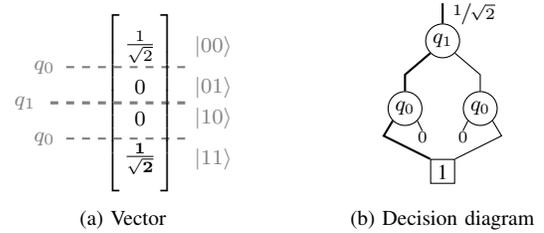

The general idea of decision diagrams~\cite{niemannQMDDsEfficientQuantum2016,zulehnerHowEfficientlyHandle2019} is about uncovering and exploiting redundancies within the involved quantum states and operations.
More precisely, consider a quantum register composed of $n$ qubits $q_{n-1}, \ldots, q_1, q_0$, where $q_{n-1}$ represents the most significant qubit.
The first $2^{n-1}$ entries of the corresponding state vector represent amplitudes for basis states where $q_{n-1}$ is $\ket{0}$ and the remaining $2^{n-1}$ entries represent amplitudes where $q_{n-1}$ is~$\ket{1}$. 
This is represented in a decision diagram by a node labeled $q_{n-1}$ connected to two successor nodes labeled $q_{n-2}$, representing the zero- and one-successor. 
This process is repeated recursively until \mbox{sub-vectors} of size~1 (i.e., individual complex numbers) remain, which are connected to terminal nodes. 

During this decomposition process, equivalent \mbox{sub-vectors} are represented by the same node---reducing the overall size of the decision diagram. Furthermore, instead of having distinct terminal nodes for all amplitudes, edge weights are used to store common factors of the amplitudes. Having encoded a state vector into a decision diagram, specific amplitudes can be reconstructed multiplying the edge weights along the corresponding path. To improve the readability of decision diagrams, edge weights of $1$ are typically omitted from the visualization and nodes with an incoming edge weight of zero are shown as $0$-stubs to indicate that the whole sub-part is zero.

\begin{example}
	\autoref{fig:strong-statevector} depicts the quantum register $ \ket{\psi^\prime} $ in both, the vector and the decision diagram representation. 
	The annotations of the state vector in \autoref{fig:strong-statevectorvector} indicate how the corresponding decision diagram is constructed. 
	In order to reconstruct specific amplitudes from the decision diagram, the edge weights of the corresponding path need to be multiplied. 
	For example, reconstructing the amplitude of the state $\ket{00}$ (bold line in the figure) requires multiplying the edge weight of the root edge (${\sfrac{1}{\sqrt{2}}}$) with the right edge of $q_1$ ($1$) as well as $q_0$ ($1$), i.e.~${\sfrac{1}{\sqrt{2}}} \cdot 1 \cdot 1 = {\sfrac{1}{\sqrt{2}}}$.
\end{example}

Decision diagram representation of matrices are constructed in an analogous fashion to vectors, decomposing the matrix recursively into quarters instead of halves.
Just as the underlying vectors and matrices, decision diagrams support multiplication and addition, enabling their usage in different design automation tasks, such as quantum circuit simulation~(e.g.,~\cite{zulehnerAdvancedSimulationQuantum2019}) or equivalence checking~(e.g.,~\cite{burgholzerAdvancedEquivalenceChecking2021}).
A \mbox{web-based} visualizing tool providing an intuition of decision diagrams is available at~\url{https://iic.jku.at/eda/research/quantum_dd/tool/}~\cite{willeVisualizingDecisionDiagrams2021}.

\section{Tensor Networks}\label{sec:tns}
Tensor networks can help alleviate the complexity of the array-based simulation by exploiting redundancies in the topological structure of the quantum circuit~\cite{fannesFinitelyCorrelatedStates1992, bridgemanHandwavingInterpretiveDance2017}. 
To translate a quantum circuit into a tensor networks, each object, be it a state or a operation, is represented by a multidimensional array of complex numbers, a \emph{tensor}, connecting to other tensors according to the underlying quantum circuit. 
The extraction of useful information from such a network then typically requires the pairwise contraction of tensors into a single remaining tensor. 

\begin{example}\label{ex:matmultensor}
	Let $A, B, C$ be matrices in $\mathbb{C}^{N\times N}$. Further, let the matrix product $C=AB$ be given by
	\begin{align*}
	{C_{i,j} = \textstyle\sum_{k=0}^{N-1} A_{i,k}B_{k,j}},
	\end{align*}
	with $i,j=0,\ldots,N-1$.
	Then, this corresponds to the contraction of the rank-$2$ tensors $A = [A_{i,k}]$ and $B = [B_{k,j}]$ over the shared index $k$. 
	This is conveniently represented graphically as:
	\begingroup\centering
		\begin{tikzpicture}[scale=1.2, every node/.style={scale=1.2}]
			\node[draw] (C) {$C$};
			\node[draw, right = 3 of C] (A) {$A$};			
			\node[draw, right = 1 of A] (B) {$B$};
			\draw[] (C) -- ++ (-0.5, 0) node[left] {$i$};
			\draw[] (C) -- ++ (0.5, 0) node[right] {$j$};
			\draw[] (A) -- ++ (-0.5, 0) node[left] {$i$};
			\draw[] (B) -- ++ (0.5, 0) node[right] {$j$};
			\draw[] (A) -- (B) node[above, midway] {$k$};
			\node[] at ($(C)!0.5!(A)$) {$=$};			
		\end{tikzpicture}
	\endgroup
\end{example}

The order in which all the tensors are contracted is called \emph{contraction plan}. 
The main goal of such a plan is to keep the intermediate tensors and their dimension of contracted indices (also referred to as \emph{bond dimension}) during the computation in check---a task  proven to be NP-hard~\cite{chi-chungOptimizingClassMultidimensional1997}. 
Therefore, a plethora of methods have been developed to efficiently determine suitable contraction plans~\cite{grayHyperoptimizedTensorNetwork2021}.

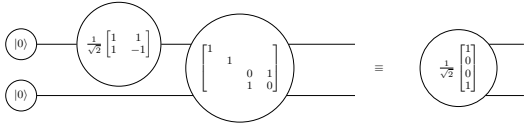
\begin{figure}
\centering
\resizebox{0.8\linewidth}{!}{
		\begin{tikzpicture}
            \node[circle, draw] (q2) {$\ket{0}$};
			\node[circle, draw, below = 0.5 of q2] (q1) {$\ket{0}$};
			\node[circle, draw, right = 1 of q2] (H) {$\frac{1}{\sqrt{2}}\begin{bmatrix}1 & 1\\ 1 & -1\end{bmatrix}$};
			\node (result) at (11,-0.6) [circle, draw]  {$\frac{1}{\sqrt{2}} \begin{bmatrix} 1 \\ 0 \\ 0 \\ 1 \end{bmatrix}$};
			\node (CXA) at (5.5,-0.6) [circle, draw]  {$\begin{bmatrix}1 & & &\\ & 1 & & \\ & & 0 & 1\\ & & 1 & 0\end{bmatrix}$};
			\node [right = 8 of q2] (q2o) {};
			\node [right = 8 of q1] (q1o) {};
			\node [right = 4 of q2o] (q2f) {};
			\node [right = 4 of q1o] (q1f) {};
			\draw (q2) -- (H);
			\coordinate(cpa) at (intersection 1 of CXA and q2o--H);
			\draw (H) -- (cpa);
			\coordinate(cpb) at (intersection 1 of CXA and H--q2o);
			\draw (cpb) -- (q2o);
			
			\coordinate(cpc) at (intersection 1 of CXA and q1o--q1);
			\draw (q1) -- (cpc);			
			\coordinate(cpe) at (intersection 1 of CXA and q1--q1o);
			\draw (cpe) -- (q1o);
			\node  (EQ) at (9,-0.6) {
				$\equiv$
			};
			
            \coordinate(cfin1) at (intersection 1 of result and q1--q1f);
			\draw (cfin1) -- (q1f);
			\coordinate(cfin2) at (intersection 1 of result and q2--q2f);
			\draw (cfin2) -- (q2f);

		\end{tikzpicture}}
	\caption{Tensor network representation of the quantum circuit to create the Bell state}
	\label{fig:ghz_circuit_tn}
	\vspace{-1em}
\end{figure}

\begin{example}\label{ex:circuit_tn}
Consider again the Bell state from \autoref{fig:strong-statevectorvector}. 
\autoref{fig:ghz_circuit_tn} shows how this translates to a tensor network.
Each individual tensor is illustrated by a \enquote{bubble} containing the actual data of the tensor. This representation only requires a linear amount of memory with regard to the total number of qubits and gates (in contrast to the exponential representation in the array-based method). 
The final state vector, on the other hand, still is of size $2^n$, where $n$ denotes the number of qubits in the system.
\end{example}

As shown by the example, the computation of the complete output state vector with tensor networks is generally infeasible. 
Different specialized types of tensor networks have been proposed to alleviate that complexity by imposing certain structures by decomposing the whole state into smaller tensors (see~\cite{RevModPhys.93.045003} and the references therein).

This is used, e.g., in classical quantum circuit simulation, where it is desirable to determine a single scalar quantity, such as the expected value of some observable or an individual amplitude. 
Methods based on tensor networks accomplish this by fixing the output indices of the circuit's tensor network, i.e., adding \enquote{bubbles} at the end of the circuit. 
Contracting this network results in a single rank-$0$ tensor---a scalar. 
Whenever the size and bond dimension of intermediate tensors can be kept in check, this can be done extremely efficient.

\section{ZX-Calculus}\label{sec:zx}

\begin{figure}[t]
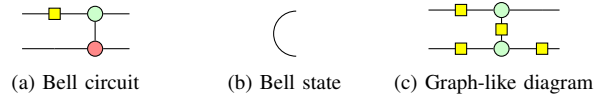

    \begin{subfigure}[b]{0.3\linewidth}
    \centering
        \begin{ZX}[ampersand replacement=\&]
    \zxN{} \rar \& [\zxWCol] \zxH{} \rar \& [\zxWCol] \zxZ{}\dar\rar \& [\zxWCol] \zxN\\[\zxHRow]
    \zxN{} \rar \& [\zxWCol] \zxN{} \rar \& [\zxWCol] \zxX{}    \rar \& [\zxWCol] \zxN{}
    \end{ZX} 
    \caption{Bell circuit}
    \label{fig:zx_bell_circuit}
    \end{subfigure}
   \begin{subfigure}[b]{0.3\linewidth}
   \centering
     \begin{ZX}[ampersand replacement=\&]
    \zxN{} \&  \zxN{}\ar[dd,C]\\[\zxWRow]
    \\[\zxHRow]
    \zxN{} \&  \zxN{}
    \end{ZX}
   \caption{Bell state}
    \label{fig:zx_bell_state}
   \end{subfigure}
   \begin{subfigure}[b]{0.3\linewidth}
    \centering
        \begin{ZX}[ampersand replacement=\&]
    \zxN{} \rar \& [\zxWCol] \zxH{} \rar \& [\zxWCol] \zxZ{}\ar[dd,H]\rar \& [\zxWCol] \zxN{} \rar \& \zxN{}\\
    \\
    \zxN{} \rar \& [\zxWCol] \zxH{} \rar \& [\zxWCol] \zxZ{}    \rar \& [\zxWCol] \zxH{} \rar \&  \zxN{}
    \end{ZX} 
    \caption{Graph-like diagram}
    \label{fig:zx_graph_like}
    \end{subfigure}
   \caption{ZX-diagrams for the Bell state}
   \label{fig:zx_diagrams}
   \vspace{-1em}
\end{figure}

The ZX-calculus~\cite{vandeweteringZXcalculusWorkingQuantum2020,coeckePicturingQuantumProcesses2018} is a graphical notation for quantum circuits equipped with a powerful set of rewrite rules that enable diagrammatic reasoning about quantum computing. A ZX-diagram is made up of colored nodes (called \emph{spiders}) that are connected by wires. Each spider can either be green
(Z-spider
\begin{ZX}
  \zxZ{}
\end{ZX}
) or red (X-spider
\begin{ZX}
  \zxX{}
\end{ZX}
) and is optionally attributed a scalar phase. Spiders without inputs are called \emph{states}, whereas spiders with no outputs are called \emph{effects}. 

An important concept for ZX-diagrams is the \emph{only connectivity matters} paradigm, which expresses the fact that two \mbox{ZX-diagrams} are considered equal if one can be transformed into the other simply by bending wires.

\begin{example}\label{ex:zx}
    Consider the Bell circuit in \autoref{fig:zx_bell_circuit}. It is equivalent to the ZX-diagram \begin{ZX}[ampersand replacement=\&]
    \zxN{} \ar[dr,s]  \& [\zxWCol] \zxN{} \rar \& [\zxWCol] \zxX{}\dar\ar[dr,s] \& [\zxWCol] \zxN{}\\
    \zxN{} \ar[ur, s] \& [\zxWCol] \zxH{} \rar \& [\zxWCol] \zxZ{}    \ar[ur,s] \& [\zxWCol] \zxN{}
    \end{ZX} because they can be transformed into each other by (un-)crossing the wires. Here, the \emph{Hadamard box} \begin{ZX}
      \zxN{} \rar & \zxH{} \rar & \zxN{}
    \end{ZX} is a short notation for the ZX-diagram \begin{ZX}[ampersand replacement=\&]
        \zxN{} \rar \& \zxFracZ{\pi}{2} \rar \& \zxFracX{\pi}{2} \rar \& \zxFracZ{\pi}{2} \rar \& \zxN{}
      \end{ZX} and represents the Hadamard transformation.
    
    To see how this circuit acts on the $\begin{ZX}
      \zxX{} \rar & \zxN{}
    \end{ZX} = \ket{0}$ states, we can plug them into the ZX-diagram and simplify with the \mbox{ZX-calculus}:
    \begin{align*}
       \begin{ZX}[ampersand replacement=\&]
    \zxX{} \rar \&  \zxH{} \rar \&  \zxZ{}\dar\rar \&  \zxN{}\\
    \zxX{} \rar \&  \zxN{} \rar \&  \zxX{}    \rar \&  \zxN{}
    \end{ZX} = 
    \begin{ZX}[ampersand replacement=\&]
    \zxZ{} \rar \&   \zxZ{}\dar\rar \&  \zxN{}\\
    \zxX{} \rar \&   \zxX{}    \rar \&  \zxN{}
    \end{ZX} =
     \begin{ZX}[ampersand replacement=\&]
    \zxZ{}\dar\rar \&  \zxN{}\\
    \zxX{}    \rar \&  \zxN{}
    \end{ZX} = 
     \begin{ZX}[ampersand replacement=\&]
    \zxN{} \&  \zxN{}\dar[C]\\[\zxWRow]
    \zxN{} \&  \zxN{}
    \end{ZX}
   \end{align*}
   At the end, the Bell state shown in \autoref{fig:zx_bell_state} is obtained.
\end{example}

Any quantum circuit can be interpreted as a ZX-diagram. ZX-diagrams are more general than quantum circuits however, and allow for representations that do not have meaningful interpretations as quantum circuits. It is this flexibility of being able to leave the quantum circuit formalism that makes the ZX-calculus a good intermediate language when working with quantum circuits. 

While the basic axiomatization of the ZX-calculus is a powerful language for quantum information theory, it is hard to apply directly in automated reasoning. The reason for this is the lack of normal-forms for ZX-diagrams---an important feature for automated rewriting. The backbone of many automated methods using the ZX-calculus is an alternate representation of \mbox{ZX-diagrams} using only Z-spiders and wires with Hadamard boxes, called \emph{graph-like} ZX-diagrams. The graph-like \mbox{ZX-diagram} corresponding to the Bell circuit is shown in \autoref{fig:zx_graph_like}. Additional rewrite rules based on graph-theoretic simplification are defined for these graph-like diagrams enabling the definition of a terminating rewriting procedure~\cite{Duncan2020graphtheoretic}.

This automated rewriting as well as the ability to efficiently work with quantum operations such as \emph{phase polynomials} has led to many algorithms based on the ZX-calculus being used to solve problems in the design automation of quantum circuits such as compilation and optimization as well as simulation and verification~\cite{kissingerReducingTcountZXcalculus2020, 10.1088/2058-9565/ac5d20,debeaudrapTechniquesReducePi2020,cowtanGenericCompilationStrategy2020}.

\section{Conclusion}
\label{sec:conclusion}

In this work, we briefly reviewed the basics of arrays, decision diagrams, tensor networks, and ZX-calculus as well as their applications in design automation for quantum computing.
Each of these complementary data structures provides a certain trade-off between memory consumption, performance, and conceptional complexity.
Picking the most suitable data structure for the job at hand is crucial to have an efficient workflow in design automation for quantum computing.
We hope this work provides the interested reader with an intuition on the data structures and necessary pointers to further explore the wide range of possible methods and applications.

\section*{Acknowledgements}
{\small 
This project has received funding from the European Research Council (ERC) under the European Union’s Horizon 2020 research and innovation programme (grant agreement No. 101001318).
It is part of the Munich Quantum Valley, which is supported by the Bavarian state government with funds from the Hightech Agenda Bayern Plus and was partially supported by the BMK, BMDW, and the State of Upper Austria in the frame of the COMET program (managed by the FFG).
}

\printbibliography

@inproceedings{amyLargescaleFunctionalVerification2019,
  title = {Towards large-scale functional verification of universal quantum circuits},
  booktitle = {International {{Conference}} on {{Quantum Physics}} and {{Logic}}},
  author = {Amy, Matthew},
  date = {2019-01},
  eventtitle = {International {{Conference}} on {{Quantum Physics}} and {{Logic}}}
}

@inproceedings{boteaComplexityQuantumCircuit2018,
  title = {On the complexity of quantum circuit compilation},
  booktitle = {Int'l {{Symp}}. on {{Combinatorial Search}}},
  author = {Botea, A. and Kishimoto, A. and Marinescu, Radu},
  date = {2018},
  eventtitle = {Int'l {{Symp}}. on {{Combinatorial Search}}}
}

@article{bridgemanHandwavingInterpretiveDance2017,
  title = {Hand-waving and {{Interpretive Dance}}: {{An Introductory Course}} on {{Tensor Networks}}},
  shorttitle = {Hand-waving and {{Interpretive Dance}}},
  author = {Bridgeman, Jacob C. and Chubb, Christopher T.},
  date = {2017-06-02},
  journaltitle = {J. Phys. A: Math. Theor.}
}

@article{burgholzerAdvancedEquivalenceChecking2021,
  title = {Advanced equivalence checking for quantum circuits},
  author = {Burgholzer, Lukas and Wille, Robert},
  date = {2021},
  journaltitle = tcad
}

@article{chi-chungOptimizingClassMultidimensional1997,
  title = {On optimizing a class of multi-dimensional loops with reduction for parallel execution},
  author = {Chi-Chung, Lam and Sadayappan, P. and Wenger, Rephael},
  date = {1997-06},
  journaltitle = {Parallel Process. Lett.}
}

@article{RevModPhys.93.045003,
  title = {Matrix product states and projected entangled pair states: {C}oncepts, symmetries, theorems},
  author = {Cirac, J. Ignacio and P\'erez-Garc\'{i}a, David and Schuch, Norbert and Verstraete, Frank},
  journal = {Rev. Mod. Phys.},
  volume = {93},
  issue = {4},
  pages = {045003},
  numpages = {65},
  year = {2021},
  publisher = {American Physical Society},
  doi = {10.1103/RevModPhys.93.045003},
}

@inproceedings{coeckePicturingQuantumProcesses2018,
  title = {Picturing quantum processes},
  booktitle = {Diagrammatic {{Representation}} and {{Inference}}},
  author = {Coecke, Bob and Kissinger, Aleks},
  editor = {Chapman, Peter and Stapleton, Gem and Moktefi, Amirouche and Perez-Kriz, Sarah and Bellucci, Francesco},
  date = {2018}
}

@misc{cordierBiologyMedicineLandscape2021,
  title = {Biology and medicine in the landscape of quantum advantages},
  author = {Cordier, Benjamin A. and Sawaya, Nicolas P. D. and Guerreschi, Gian G. and McWeeney, Shannon K.},
  date = {2021-12-16},
  eprint = {2112.00760},
  eprinttype = {arxiv},
  archiveprefix = {arXiv}
}

@misc{cowtanGenericCompilationStrategy2020,
  title = {A generic compilation strategy for the unitary coupled cluster ansatz},
  author = {Cowtan, Alexander and Simmons, Will and Duncan, Ross},
  date = {2020-07-20},
  eprint = {2007.10515},
  eprinttype = {arxiv},
  archiveprefix = {arXiv}
}

@article{debeaudrapTechniquesReducePi2020,
  title = {Techniques to reduce {$\pi/4$}-parity-phase circuits, motivated by the {{ZX}} calculus},
  author = {family=Beaudrap, given=Niel, prefix=de, useprefix=true and Bian, Xiaoning and Wang, Quanlong},
  date = {2020-05-01},
  journaltitle = {Electron. Proc. Theor. Comput. Sci.},
  volume = {318},
  pages = {131--149},
  doi = {10.4204/EPTCS.318.9}
}

@article{Duncan2020graphtheoretic,
  doi = {10.22331/q-2020-06-04-279},
  title = {Graph-theoretic {S}implification of {Q}uantum {C}ircuits with the {ZX}-calculus},
  author = {Duncan, Ross and Kissinger, Aleks and Perdrix, Simon and van de Wetering, John},
  journal = {{Quantum}},
  issn = {2521-327X},
  publisher = {{Verein zur F{\"{o}}rderung des Open Access Publizierens in den Quantenwissenschaften}},
  volume = {4},
  pages = {279},
  month = jun,
  year = {2020}
}

@article{fannesFinitelyCorrelatedStates1992,
  title = {Finitely correlated states on quantum spin chains},
  author = {Fannes, Mark and Nachtergaele, Bruno and Werner, Reinhard F},
  date = {1992},
  journaltitle = {Commun. Math. Phys.},
  volume = {144},
  number = {3},
  pages = {443--490}
}

@article{gambettaIBMRoadmapScaling2020,
  title = {{{IBM}}’s {{Roadmap For Scaling Quantum Technology}}},
  author = {Gambetta, Jay},
  date = {2020-09-15},
  journaltitle = {IBM Research Blog},
  url = {https://www.ibm.com/blogs/research/2020/09/ibm-quantum-roadmap/},
  entrysubtype = {newspaper}
}

@article{grayHyperoptimizedTensorNetwork2021,
  title = {Hyper-optimized tensor network contraction},
  author = {Gray, Johnnie and Kourtis, Stefanos},
  date = {2021-03-15},
  journaltitle = {Quantum},
  volume = {5},
  pages = {410}
}

@article{grurlnoiseaware2022,
	title = {Noise-aware Quantum Circuit Simulation With Decision Diagrams},
	author = {Thomas Grurl and Jürgen Fuß and Robert Wille},
	date = {2022}
}

@article{hanerSoftwareMethodologyCompiling2018,
  title = {A software methodology for compiling quantum programs},
  author = {Häner, Thomas and Steiger, Damian S. and Svore, Krysta and Troyer, Matthias},
  date = {2018-02},
  journaltitle = {Quantum Sci. Technol.},
  volume = {3},
  number = {2},
  pages = {020501},
  doi = {10.1088/2058-9565/aaa5cc}
}

@misc{hermanSurveyQuantumComputing2022,
  title = {A survey of quantum computing for finance},
  author = {Herman, Dylan and Googin, Cody and Liu, Xiaoyuan and Galda, Alexey and Safro, Ilya and Sun, Yue and Pistoia, Marco and Alexeev, Yuri},
  date = {2022-01-18},
  eprint = {2201.02773},
  eprinttype = {arxiv},
  archiveprefix = {arXiv}
}

@inproceedings{hillmichAccurateNeededEfficient2020,
  title = {As accurate as needed, as efficient as possible: {{Approximations}} in {{DD-based}} quantum circuit simulation},
  shorttitle = {As accurate as needed, as efficient as possible},
  booktitle = date,
  author = {Hillmich, Stefan and Kueng, Richard and Markov, Igor L. and Wille, Robert},
  date = {2020-12-10},
  eventtitle = {date}
}

@misc{hongTensorNetworkBased2020,
  title = {A tensor network based decision diagram for representation of quantum circuits},
  author = {Hong, Xin and Zhou, Xiangzhen and Li, Sanjiang and Feng, Yuan and Ying, Mingsheng},
  date = {2020-09-05},
  eprint = {2009.02618},
  eprinttype = {arxiv},
  archiveprefix = {arXiv}
}

@misc{huangProvablyEfficientMachine2022,
  title = {Provably efficient machine learning for quantum many-body problems},
  author = {Huang, Hsin-Yuan and Kueng, Richard and Torlai, Giacomo and Albert, Victor V. and Preskill, John},
  date = {2022-02-20},
  eprint = {2106.12627},
  eprinttype = {arxiv},
  archiveprefix = {arXiv}
}

@inproceedings{jonesQuESTHighPerformance2018,
  title = {{{QuEST}} and high performance simulation of quantum computers},
  booktitle = {Scientific {{Reports}}},
  author = {Jones, Tyson and Brown, Anna and Bush, Ian and Benjamin, Simon C.},
  date = {2018}
}

@article{kandalaHardwareefficientVariationalQuantum2017,
  title = {Hardware-efficient variational quantum eigensolver for small molecules and quantum magnets},
  author = {Kandala, Abhinav and Mezzacapo, Antonio and Temme, Kristan and Takita, Maika and Brink, Markus and Chow, Jerry M. and Gambetta, Jay M.},
  date = {2017-09},
  journaltitle = {Nature},
  volume = {549},
  number = {7671},
  pages = {242--246}
}

@article{kissingerReducingTcountZXcalculus2020,
  title = {Reducing {{T-count}} with the {{ZX-calculus}}},
  author = {Kissinger, Aleks and family=Wetering, given=John, prefix=van de, useprefix=true},
  date = {2020-08-11},
  journaltitle = {Phys. Rev. A}
}

@article{10.1088/2058-9565/ac5d20,
	author={Kissinger, Aleks and van de Wetering, John},
	title={Simulating quantum circuits with ZX-calculus reduced stabiliser decompositions},
	journal={Quantum Science and Technology},
	year={2022}
}

@inproceedings{liTacklingQubitMapping2019,
  title = {Tackling the qubit mapping problem for {{NISQ-era}} quantum devices},
  booktitle = {Int'l {{Conf}}. on {{Architectural Support}} for {{Programming Languages}} and {{Operating Systems}}},
  author = {Li, Gushu and Ding, Yufei and Xie, Yuan},
  date = {2019},
  eventtitle = {Int'l {{Conf}}. on {{Architectural Support}} for {{Programming Languages}} and {{Operating Systems}}}
}

@book{nielsenQuantumComputationQuantum2010,
  title = {Quantum {{Computation}} and {{Quantum Information}}},
  shorttitle = {Quantum {{Computation}} and {{Quantum Information}}},
  author = {Nielsen, Michael A. and Chuang, Isaac L.},
  date = {2010},
  publisher = {{Cambridge University Press}}
}

@inproceedings{niemannEquivalenceCheckingMultilevel2014,
  title = {Equivalence checking in multi-level quantum systems},
  booktitle = rc_conf,
  author = {Niemann, Philipp and Wille, Robert and Drechsler, Rolf},
  date = {2014},
  eventtitle = {rc\_conf}
}

@article{niemannQMDDsEfficientQuantum2016,
  title = {{{QMDDs}}: {{Efficient}} quantum function representation and manipulation},
  author = {Niemann, Philipp and Wille, Robert and Miller, David Michael and Thornton, Mitchell A. and Drechsler, Rolf},
  date = {2016},
  journaltitle = tcad
}

@article{shorPolynomialtimeAlgorithmsPrime1997,
  title = {Polynomial-time algorithms for prime factorization and discrete logarithms on a quantum computer},
  author = {Shor, Peter W.},
  date = {1997},
  journaltitle = {SIAM J. Comput.}
}

@inproceedings{smithQuantumComputationalCompiler2019,
  title = {A quantum computational compiler and design tool for technology-specific targets},
  booktitle = {Int'l {{Symp}}. on {{Computer Architecture}}},
  author = {Smith, Kaitlin N. and Thornton, Mitchell A.},
  date = {2019},
  pages = {579--588}
}

@misc{vandeweteringZXcalculusWorkingQuantum2020,
  title = {{{ZX-calculus}} for the working quantum computer scientist},
  author = {family=Wetering, given=John, prefix=van de, useprefix=true},
  date = {2020-12-27},
  eprint = {2012.13966},
  eprinttype = {arxiv},
  archiveprefix = {arXiv}
}

@inproceedings{viamontesCheckingEquivalenceQuantum2007,
  title = {Checking equivalence of quantum circuits and states},
  booktitle = iccad,
  author = {Viamontes, George F. and Markov, Igor L. and Hayes, John P.},
  date = {2007},
  eventtitle = {iccad}
}

@inproceedings{wangXQDDbasedVerificationMethod2008,
  title = {An {{XQDD-based}} verification method for quantum circuits},
  booktitle = ieice,
  author = {Wang, S.-A. and Lu, C.-Y. and Tsai, I-M. and Kuo, S.-Y.},
  date = {2008},
  pages = {584--594}
}

@inproceedings{willeVisualizingDecisionDiagrams2021,
  title = {Visualizing decision diagrams for quantum computing},
  booktitle = date,
  author = {Wille, Robert and Burgholzer, Lukas and Artner, Michael},
  date = {2021-02},
  eventtitle = {date}
}

@inproceedings{yamashitaFastEquivalencecheckingQuantum2010,
  title = {Fast equivalence-checking for quantum circuits},
  booktitle = {Int'l {{Symp}}. on {{Nanoscale Architectures}}},
  author = {Yamashita, S. and Markov, I. L.},
  date = {2010-06},
  eventtitle = {Int'l {{Symp}}. on {{Nanoscale Architectures}}}
}

@article{doi:10.1126/science.abe8770,
	author = {Han-Sen Zhong  and Hui Wang  and Yu-Hao Deng  and Ming-Cheng Chen  and Li-Chao Peng  and Yi-Han Luo  and Jian Qin  and Dian Wu  and Xing Ding  and Yi Hu  and Peng Hu  and Xiao-Yan Yang  and Wei-Jun Zhang  and Hao Li  and Yuxuan Li  and Xiao Jiang  and Lin Gan  and Guangwen Yang  and Lixing You  and Zhen Wang  and Li Li  and Nai-Le Liu  and Chao-Yang Lu  and Jian-Wei Pan },
	title = {Quantum computational advantage using photons},
	journal = {Science},
	volume = {370},
	number = {6523},
	pages = {1460-1463},
	year = {2020},
	doi = {10.1126/science.abe8770},
}

@article{zulehnerAdvancedSimulationQuantum2019,
  title = {Advanced simulation of quantum computations},
  author = {Zulehner, Alwin and Wille, Robert},
  date = {2019},
  journaltitle = tcad
}

@article{zulehnerEfficientMethodologyMapping2019,
  title = {An efficient methodology for mapping quantum circuits to the {{IBM QX}} architectures},
  author = {Zulehner, Alwin and Paler, Alexandru and Wille, Robert},
  date = {2019},
  journaltitle = tcad
}

@inproceedings{zulehnerHowEfficientlyHandle2019,
  title = {How to efficiently handle complex values? {{Implementing}} decision diagrams for quantum computing},
  shorttitle = {How to efficiently handle complex values?},
  booktitle = iccad,
  author = {Zulehner, Alwin and Hillmich, Stefan and Wille, Robert},
  date = {2019},
  eventtitle = {iccad}
}

@STRING{tcad	= {{IEEE} Trans. on {CAD} of Integrated Circuits and Systems} }

@STRING{ieice	= {IEICE Trans. Fundamentals} }

@STRING{siam	= {SIAM Jour. of Comp.} }

@STRING{iccad	= {Int'l Conf. on CAD} }

@STRING{date	= {Design, Automation and Test in Europe} }

@STRING{rc_conf	= {Int'l Conf. of Reversible Computation} }

@article{DeRaedt2019,
author = {{De Raedt}, Hans and others},
journal = {Computer Physics Communications},
pages = {47--61},
title = {Massively parallel quantum computer simulator, eleven years later},
volume = {237},
year = {2019}
}

@BOOK{VMP:2009,
  author =       {G. F. Viamontes and I. L. Markov and J. P. Hayes},
  title =        {Quantum Circuit Simulation},
  publisher =    {Springer},
  year =         {2009}
}

@article{PhysRevLett.116.250501,
  title = {Improved Classical Simulation of Quantum Circuits Dominated by {Clifford} Gates},
  author = {Bravyi, Sergey and Gosset, David},
  journal = {Phys. Rev. Lett.},
  volume = {116},
  issue = {25},
  pages = {250501},
  numpages = {5},
  year = {2016},
  publisher = {American Physical Society},
  doi = {10.1103/PhysRevLett.116.250501}
}
\end{document}